# Edsger Dijkstra

## The Man Who Carried Computer Science on His Shoulders

### *Krzysztof Apt*

As it turned out, the train I had taken from Nijmegen to Eindhoven arrived late. To make matters worse, I was then unable to find the right office in the university building. When I eventually arrived for my appointment, I was more than half an hour behind schedule. The professor completely ignored my profuse apologies and proceeded to take a full hour for the meeting. It was the first time I met Edsger Wybe Dijkstra.

At the time of our meeting in 1975, Dijkstra was 45 years old. The most prestigious award in computer science, the ACM Turing Award, had been conferred on him three years earlier. Almost twenty years his junior, I knew very little about the field—I had only learned what a flowchart was a couple of weeks earlier. I was a postdoc newly arrived from communist Poland with a background in mathematical logic and a plan to stay in the West. I left the meeting with two book recommendations and a copy of a recent research article Dijkstra had written. He also suggested that I learn the programming language Pascal.

Dijkstra passed away in 2002. During the 1970s and 1980s, at the height of his career, he was probably the most discussed scientist in his field. He was a pioneer and a genius whose work and ideas shaped the emerging field of computer science like few others. It was over the course of his career that computer science became a respectable and established discipline.

There is an enduring tension between the engineering view of computer science, which is focused on building software systems and hardware components, and the mathematical and logical view, which aims to provide rigorous foundations for areas such as programming. Dijkstra tried to reconcile both views. As a result, he contributed to both sides of the divide in a number of fundamental ways.

Dijkstra was also a most striking and unusual person. He was admired and criticized, in equal measure, and commented upon by almost everyone he came into contact with. Yet, despite his achievements, Dijkstra has always remained largely unknown outside computer science. Eighteen years after his death, few people have heard of him, even in his own country.

Edsger Dijkstra was born in Rotterdam in 1930. He described his father, at one time the President of the Dutch Chemical Society, as "an excellent chemist," and his mother as "a brilliant mathematician who had no job."[1] In 1948, Dijkstra achieved remarkable results when he completed secondary school at the famous *Erasmiaans Gymnasium* in Rotterdam. His school diploma shows that he earned the highest possible grade in no less than six out of thirteen subjects. He then enrolled at the University of Leiden to study physics.

In September 1951, Dijkstra's father suggested he attend a three-week course on programming in Cambridge. It turned out to be an idea with far-reaching consequences. It was in Cambridge that Dijkstra met the mathematician and computer scientist Adriaan van Wijngaarden, who subsequently offered him a job at the Mathematisch Centrum (Mathematical Centre) in Amsterdam, which he joined the following year. Dijkstra became, in his own words, "the first Dutchman with the qualification 'programmer' on the payroll."[2] In 1956, he finished his studies in Leiden. Three years later, he defended his PhD thesis, "Communication with an Automatic Computer." His supervisor was van Wijngaarden.

Dijkstra worked at the Mathematisch Centrum until 1962, when he moved to Eindhoven to assume the position of Professor in the Mathematics Department of the Eindhoven University of Technology. In 1973, he reduced his employment to one day a week and for the remaining four days worked as a research fellow at the Burroughs Corporation, at that time an innovative American computer manufacturer. His only duties for Burroughs involved undertaking research and traveling to the US a few times each year to visit the company headquarters.

In Dijkstra's reports, he listed the address Plataanstraat 5, Nuenen 4565, The Netherlands. This led some to assume that the Burroughs Corporation had opened a new office. The address was, in fact, that of Dijkstra's home, a modest house situated in a village near the outskirts of Eindhoven. His office consisted of a small room on the second floor, which was equipped with an "elegant portable Olivetti typewriter" and "two telephones ... one of which he could use to call anywhere in the world, with the bills going direct to Burroughs."[3]

In 1984, disenchanted with a change of direction at the Burroughs Corporation and a lack of support for computer science at his university, Dijkstra left the Netherlands and took up a prestigious chair in computer science at the University of Texas at Austin. "Whereas Burroughs's intellectual horizon was shrinking," he later wrote, "my own was widening."[4] He retired in 1999.

In early 2002, Dijkstra learned that he was terminally ill and moved back to Nuenen with his wife, Ria. He passed away in August, just a few months after returning. Ria died ten years later. The couple are survived by three children: Femke, Marcus, and Rutger.

Over the course of his career, Dijkstra wrote around 40 journal publications and 30 conference publications.[5] He is listed as the sole author for almost all of these works. Several of his journal papers were just a couple of pages long, while most of his conference publications were non-refereed manuscripts that he presented during the Annual International Marktoberdorf Summer School and published in the school proceedings. He also wrote a handful of book chapters and a few books.

Viewed from this perspective, Dijkstra's research output appears respectable, but otherwise unremarkable by current standards. In this case, appearances are indeed deceptive. Judging his body of work in this manner misses the mark completely. Dijkstra was, in fact, a highly prolific writer, albeit in an unusual way.

I N 1959, DIJKSTRA BEGAN writing a series of private reports. Consecutively numbered and with his initials as a prefix, they became known as EWDs. He continued writing these reports for more than forty years. The final EWD, number 1,318, is dated April 14, 2002. In total, the EWDs amount to over 7,700 pages. Each report was photocopied by Dijkstra himself and mailed to other computer scientists. The recipients varied depending on the subject. Around 20 copies of each EWD were distributed in this manner.

The EWDs were initially written in Dutch using a typewriter. In 1972, Dijkstra switched to writing exclusively in English, and in 1979 he began writing them mostly by hand. The EWDs consisted of research papers, proofs of new or existing theorems, comments or opinions on the scientific work of others (usually critical and occasionally titled "A somewhat open letter to..."), position papers, transcripts of speeches, suggestions on how to conduct research ("Do only what only you can do"), opinions on the role of education and universities ("It is not the task of the University to offer what society asks for, but to give what society needs"[6]), and original solutions to puzzles. Later reports also included occasional accounts of Dijkstra's life and work. A number of EWDs are titled "Trip Report" and provide detailed descriptions of his travels to conferences ("I managed to visit Moscow without being dragged to the Kremlin"[7]), summer schools, or vacation destinations. These reports are a rich source of information about Dijkstra's habits, views, thinking, and (hand) writing. Only a small portion of the EWDs concerned with research ever appeared in scientific journals or books.

This way of reporting research was, in fact, common during the eighteenth century. In the twentieth century it was a disarming anachronism. Nevertheless, it worked. In EWD1000, dated January 11, 1987, Dijkstra recounts being told by readers that they possessed a sixth or seventh generation copy of EWD249.[8]

Whether written using a fountain pen or typewriter, Dijkstra's technical reports were composed at a speed of around three words per minute. "The rest of the time," he remarked, "is taken up by thinking."[9] For Dijkstra, writing and thinking blended into one activity. When preparing a new EWD, he always sought to produce the final version from the outset.

Around 1989, Hamilton Richards, a former colleague of Dijkstra's in Austin, created a website to preserve all the available EWDs and their bibliographic entries.[10] The E. W. Dijkstra Archive, as the site is known, also offers an abundance of additional material about Dijkstra, including links to scans of his early technical reports, interviews, videos, obituaries, articles, and a blog.

Despite a worldwide search, a number of EWDs from the period prior to 1968 have never been found. Other missing entries in the numbering scheme were, by Dijkstra's own admission, "occupied by documents that I failed to complete."[11]

D IJKSTRA WAS A true pioneer in his field. This occasionally caused him problems in everyday life. In his Turing Award lecture he recalled:

In 1957, I married, and Dutch marriage rites require you to state your profession and I stated that I was a programmer. But the municipal authorities of the town of Amsterdam did not accept it on the grounds that there was no such profession.[12]

In the mid-1950s, Dijkstra conceived an elegant shortest path algorithm. There were very few computer science journals at the time and finding somewhere to publish his three-page report proved far from easy. Eventually, three years later, he settled on the newly established *Numerische Mathematik*.[13] "A Note on Two Problems in Connexion with Graphs" remains one of the most highly cited papers in computer science, while Dijkstra's algorithm has become indispensable in GPS navigation systems for computing the shortest route between two locations.

Over a period of eight months beginning in December 1959, Dijkstra wrote an ALGOL 60 compiler with Jaap Zonneveld.[14] Theirs was the first compiler for this new and highly innovative programming language. It was a remarkable achievement. In order to write the compiler, several challenges had to be overcome. The most obvious problem the pair faced was that the machine designated to run the software, the Dutch Electrologica X1 computer, had a memory of only 4,096 words. By comparison, the memory of a present-day laptop is larger by a factor of a million.

The programming language itself was not without its own challenges. ALGOL 60 included a series of novel features, such as recursion, which was supported in a much more complex manner than logicians had ever envisaged.[15] One of the ideas suggested by Dijkstra, termed a display, addressed the implementation of recursive procedures and has since become a standard technique in compiler writing.[16]

ALGOL 60 was designed by an international committee. Although Dijkstra attended several meetings during the design process, his name does not appear among the thirteen editors of the final report.[17] Apparently, he disagreed with a number of majority opinions and withdrew from the committee. This was perhaps the first public sign of his fiercely held independence.

During his employment at the Eindhoven University of Technology, Dijkstra and his group wrote an operating system for the Electrologica X8, the successor to the X1. The system they created, known as the THE multiprogramming system (THE is an abbreviation of Technische Hogeschool Eindhoven), had an innovative layered functional structure, in which the higher layers depended only on the lower ones.[18]

It was during his work on this system that Dijkstra's interests began shifting to parallel programs, of which THE is an early example. These programs consist of a collection of components, each of which are traditional programs, executed in parallel. Such programs are notoriously difficult to write and analyze because they need to work correctly no matter the execution speeds of their components. Parallel programs also need to synchronize their actions to ensure exclusive access to resources. If several print jobs are dispatched at the same time by the users of a shared computer network, this should not lead to pages from the different print jobs becoming interspersed. Adding to the complexity, the components of parallel programs should not become deadlocked, waiting indefinitely for one another.

In the early 1960s, these problems had not yet been properly examined or analyzed, nor had any techniques been developed to verify potential solutions. Dijkstra identified a crucial synchronization problem, which he named the mutual exclusion problem, and published his solution in a single-page paper.[19] This work was taken from EWD123, an extensive 87-page report titled "Cooperating Sequential Processes." In the same report, he introduced the first known synchronization primitive, which he termed a semaphore, that led to a much simpler solution to the mutual exclusion problem.[20] He also identified the deadlock problem, which he named the deadly embrace, and proposed an elegant solution, the banker's algorithm.[21] The mutual exclusion problem, along with deadlock detection and prevention, are now mandatory topics in courses on operating systems and parallel programming.

In 1968, Dijkstra published a two-page letter addressed to the editor of the *Communications of the ACM*, in which he critiqued the goto programming statement.[22] Entitled "Go To Statement Considered Harmful," the letter was widely criticized and generated considerable debate. In the end, Dijkstra's views prevailed. Every programmer is now aware that using the goto statement leads to so-called spaghetti code. Java, currently one of the most widely used programming languages, was originally released in 1996 and does not have the goto statement. The phrase "considered harmful" is still used often in computer science and remains inextricably associated with Dijkstra.

In 1968, Dijkstra suffered a long, deep depression that persisted for almost half a year. He later made mention of being hospitalized during this period.[23] One reason for Dijkstra's torment was that his department did not consider computer science important and disbanded his group. He also had to decide what to work on next. Dijkstra's major software projects, the ALGOL 60 compiler and the THE multiprogramming system, had given him a sense that programming was an activity with its own rules. He then attempted to discover those rules and present them in a meaningful way. Above all, he strove to transform programming into a mathematical discipline, an endeavor that kept him busy for several years to come. At the time, these were completely uncharted waters. Nobody else seemed to be devoting their attention to such matters.

A year later, the appearance of the 87-page EWD249, "Notes on Structured Programming," marked the end of Dijkstra's depression.[24] The subject of the EWD was so novel, the writing so engaging, and the new term "structured programming" so convincing that the report became a huge success. But, in Dijkstra's view, "IBM ... stole the term 'Structured Programming' and ... trivialized the original concept to the abolishment of the goto statement."[25] The claim was unsurprising to those aware of Dijkstra's long-held and largely negative views toward IBM computers and software. Nowadays it is not uncommon to see similar criticisms of large corporations, but in the 1970s and 1980s few academics were prepared to take a public stand against computer manufacturers.

In 1972, Dijkstra received the ACM Turing Award, widely considered the most important prize in computer science. He was recognized for

> fundamental contributions to programming as a high, intellectual challenge; for eloquent insistence and practical demonstration that programs should be composed correctly, not just debugged into correctness; for illuminating perception of problems at the foundations of program design.[26]

Further fundamental contributions were to follow. In 1974, Dijkstra published a two-page article in which he introduced a new concept: self-stabilization.[27] In the paper, he posed the problem of how a system of communicating machines might repair itself when a temporary fault arises in one of the machines. He presented new protocols that guaranteed correct functioning of the system would eventually be restored. "Self-stabilization," he remarked, "... could be of relevance on a scale ranging from *a worldwide network* [emphasis added] to common bus control." This is a striking observation when one considers that the World Wide Web was developed just 15 years later. As it turned out, the paper was completely ignored until 1983, when Leslie Lamport stressed its importance in an invited talk. In time, the ideas outlined by Dijkstra would lead to the emergence of a whole new area in distributed computing with its own annual workshops and conferences. In 2002, the paper won an award that was posthumously renamed the Edsger W. Dijkstra Prize in Distributed Computing.

The notion that some events cannot be deterministically predicted, usually referred to as indeterminism, has kept philosophers, and later physicists, occupied for centuries. Computer scientists enter the story more recently, studying the idea under the name nondeterminism—not a reference, it should be noted, to any probabilistic interpretation of events. In 1963, John McCarthy introduced nondeterminism in the context of programming languages. A couple of years later, Robert Floyd showed how this concept, now known as angelic nondeterminism, can substantially simplify programming tasks requiring a search.[28] When choices arise there is *some* computation that delivers the desired result—though it is not certain which one.

Dijkstra's view of nondeterminism was likely influenced by the inherently nondeterministic behavior of the real-time interrupt handler he developed in his PhD thesis. In a 1975 paper, he introduced a small programming language that he called guarded commands; it encapsulated what is now termed demonic nondeterminism.[29] This was, in fact, the paper he handed me the first time I met him. In contrast to the angelic variant, *all* computations have to deliver the desired result. This more demanding view of nondeterminism—referred to as *nondeterminacy* by Dijkstra—sometimes yields simpler programs, but for reasons other than angelic nondeterminism. The programmer is free to leave some decisions unspecified.

The programming notation introduced by Dijkstra occasionally leads to elegant programs. He illustrated this point by reconsidering Euclid's 2,300-year-old algorithm for computing the greatest common divisor of two natural numbers. The algorithm can be stated as follows: As long as the two numbers differ, keep subtracting the smaller number from the bigger one. In Dijkstra's language this algorithm can be written in a simple manner that is not far removed from its description in English.[30] His language also introduced the crucial notion of a *guard*, which has since become a natural concept in various programming formalisms. Similarly, *weakest precondition semantics*, a concept Dijkstra introduced to describe program semantics, marked a late but highly significant entry into the field of program verification. A couple of years later, the language was generalized by Tony Hoare to create a highly influential programming language proposal for distributed computing that he named CSP.[31]

Dijkstra's landmark book *A Discipline of Programming* was published in 1976.[32] It introduced a novel approach to programming in which Dijkstra combined weakest precondition semantics with a number of heuristics to develop several computer programs, hand in hand with their correctness proofs. In contrast with EWD249, "Notes on Structured Programming," he was now arguing about program correctness in a formal way. This development marked a new stage in Dijkstra's research. He now viewed the development of a correct program as the development of a mathematical proof, something to which he first alluded in 1973 as part of EWD361, "Programming as a Discipline of Mathematical Nature."[33] This methodology was soon employed by Dijkstra and a group of researchers to systematically derive various, usually small, nontrivial programs. In contrast to some of his other innovations, it never really caught on.

I N THE EARLY 1980s, Dijkstra co-wrote two short but influential papers in which he applied his methodology to the systematic development of distributed programs.[34] He also sought to have this approach to programming taught to first-year students, and, with this goal in mind, put together an elegant introductory textbook with Wim Feijen.[35]

Dijkstra's realization that programming could be viewed as a mathematical activity led to his interest in analyzing mathematical reasoning. He attempted to come up with guidelines and heuristics that facilitated the discovery of proofs. In a number of cases these principles pointed him toward interesting generalizations of known results that had somehow eluded others.

A good example is the Pythagorean theorem, which is taught at secondary schools. The theorem states that in a right-angled triangle the square of the hypotenuse, *c*, is equal to the sum of the squares of the other two sides, *a* and *b*.

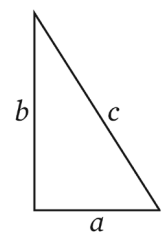

$$a^2 + b^2 = c^2$$

In 1940, Elisha Loomis collected no less than 370 proofs in *The Pythagorean Proposition*, starting with the proof that appeared in Euclid's Elements, written around 300 BCE.[36] New proofs occasionally appear to this day.

In 1986, Dijkstra came up with the following generalization to arbitrary triangles, which he included in EWD975:

> Consider a triangle with the side lengths $a$, $b$ and $c$ and the angles $\alpha$, $\beta$ and $\gamma$, lying opposite a, b and c. Then the signs of the expressions $a^2 + b^2 - c^2$ and $\alpha + \beta - \gamma$ are the same (that is, they are either both positive or both zero or both negative).

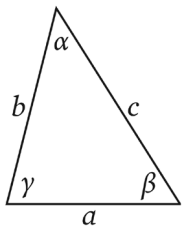

$$\text{sign}(a^2 + b^2 - c^2) = \text{sign}(\alpha + \beta - \gamma)$$

A mathematician might observe that this all seems quite obvious. Yet, apparently, nobody had thought of this generalization before Dijkstra. He concluded his report by observing that it was unclear where he might publish this result. In his view, it should be taught at schools instead of the original theorem. In 2009, EWD975 was republished posthumously by *Nieuw Archief voor Wiskunde* (New Archive for Mathematics), the magazine of the Royal Dutch Mathematical Association.[37] The five-page report was reproduced in its original handwritten form.

In a 1985 lecture, "On Anthropomorphism in Science," delivered at the University of Texas at Austin's Philosophy Department, Dijkstra speculated that mathematicians stuck to the use of implication because they associated it with cause and effect. "Somehow," he observed, "in the implication 'if $A$ then $B$,' the antecedent $A$ is associated with the cause and the consequent $B$ with the effect."[38] He claimed that equivalence should be preferred over implication. This simple principle had led to his generalization of the Pythagorean theorem.

Dijkstra also attempted to apply his methodology for developing correct programs to systematically develop proofs of mathematical theorems. In EWD1016, "A Computing Scientist's Approach to a Once-Deep Theorem of Sylvester's," he derived an elegant proof of the following theorem, first conjectured in 1893 and proved 40 years later: "Consider a finite number of distinct points in the real Euclidean plane; these points are collinear or there exists a straight line through exactly 2 of them."[39]

Another example from this period is his approach to the problem of a fair coin. A coin toss is used to determine one of two outcomes in a fair way. But how can a fair outcome be achieved when the coin is biased? In 1951, John von Neumann came up with a simple solution.[40] A number of researchers then tried to figure out how to make it more efficient. Dijkstra first learned of the problem during a lecture in 1989. He solved it immediately and a little while later came up with the solution to a related problem that apparently nobody had thought of before. His solution to the related problem can be found in EWD1071, "Making a Fair Roulette from a Possibly Biased Coin." Dijkstra's modification of von Neumann's solution was not immediately obvious and relies on a classic result from number theory, Fermat's little theorem. For a change, Dijkstra submitted this article to a journal and it was published the following year as a one-page note.[41]

The development of a natural and readable notation for representing proofs and calculations was almost as important for Dijkstra as the problems under consideration. The notation he came up with forces the author not to commit what he described as "any sins of omission":

$$
\begin{aligned}
&A \\
={} &\{ \text{ hint why } A = B \} \\
&B \\
={} &\{ \text{ hint why } B = C \} \\
&C
\end{aligned}
$$

In his final EWD,[42] for example, Dijkstra explains that for $s = (a + b + c)/2$ the equality $s(s - b)(s - c) + s(s - c)(s - a) = s(s - c)c$ holds, by writing out his argument as:

$$
\begin{aligned}
&s(s - b)(s - c) + s(s - c)(s - a) \\
={} &\{ \text{ algebra } \} \\
&s(s - c)(2s - a - b) \\
={} &\{ \text{ definition of } s \} \\
&s(s - c)c
\end{aligned}
$$

He used this notation in his own publications, notably in a book he wrote with his longstanding friend and colleague Carel Scholten.[43] The notation was adopted by several of his colleagues but did not spread further. EWD1300, "The Notational Conventions I Adopted, and Why," was republished posthumously and offers a unique insight into Dijkstra's extensive work on notation, a subject that kept him busy throughout his career.[44]

In the late 1980s, Dijkstra's research was described on his departmental homepage as follows: "My area of interest focuses on the streamlining of the mathematical argument so as to increase our powers of reasoning, in particular, by the use of formal techniques." It was also the subject of his course for computer science students.

During the period 1987–1990, I was a fellow faculty member in Austin and followed his course for a semester. Dijkstra invariably arrived early for class so that he could write out an unusual quotation on the blackboard. The lectures themselves were always meticulously pre-pared and usually devoted to presenting proofs of simple mathematical results. He delivered the lectures without notes, requiring only a blackboard and a piece of chalk. At the end of each lecture he would assign an elementary but nontrivial mathematical problem as homework and collect all the solutions at the following lecture. A week later he would return all the solutions with detailed comments, sometimes longer than the actual submissions, and then present his own solution in detail, stressing the presentation and use of notation. My own solutions fared reasonably well by Dijkstra's standards and were usually returned with only short comments, such as "Many sins of omission." It was, in fact, a course in orderly mathematical thinking, and nobody seemed at all bothered that it had nothing to do with computer science.

Dijkstra was a highly engaging lecturer. He knew how to captivate an audience with dramatic pauses, well-conceived remarks, and striking turns of phrase. The bigger the audience, the better he performed. While I was working in Austin I helped organize a departmental event with him as the main speaker. About two hundred people came along, some having flown in from Houston and neighboring states to attend the event. Dijkstra stole the show and delivered a mesmerizing presentation on Sylvester's theorem.

I N 1990, DIJKSTRA'S sixtieth birthday was celebrated in Austin with a large banquet featuring a distinguished group of guests, including numerous important figures in computer science. A festschrift was published by Springer-Verlag to mark the occasion. The volume began on page 0, in deference to the way Dijkstra numbered the pages of his EWDs. He took the trouble of thanking each of the 61 contributing authors by a handwritten letter.

The period that followed was marked by a visible change in Dijkstra's attitude and approach to his work. In the remaining twelve years of his life, despite producing about 250 new EWDs, he published almost nothing. These reports simply may not have met his standards for a journal publication. Many of the EWDs were devoted to systematically deriving proofs of tricky results, such as a problem from the International Mathematical Olympiad. He also used his methodology to obtain elegant solutions for classical puzzles, such as the knight's tour or the wolf, goat, and cabbage puzzle.[45] Some of the EWDs from this period contained accounts and assessments of his early contributions.

Following Dijkstra's retirement from teaching in the fall of 1999, a symposium was organized in May 2000 to honor his seventieth birthday. The event was called "In Pursuit of Simplicity" and included guest contributors from both Europe and the US. At this time, I was working at the Centre for Mathematics and Informatics (CWI) in Amsterdam and during the symposium I invited Dijkstra to give a lecture. CWI was, in fact, the new name for the Mathematisch Centrum where he had worked at the beginning of his career. Six months later Dijkstra accepted the invitation. He had never seen the new and larger building where the CWI had relocated in the early 1980s, and was deeply moved.

Prior to the lecture, the CWI's communication department issued a press release. News of the event caught the attention of a major Dutch newspaper. A journalist was dispatched and an extensive article with a prominent photo of Dijkstra was soon published. VPRO, an independent Dutch public broadcasting company, subsequently became interested in Dijkstra and sent a crew to Austin to make a half-hour-long program about him. "*Denken als discipline*" (Thinking as a Discipline) was broadcast in April 2001 as an episode of the science show *Noorderlicht* (Northern Lights).[46] The episode received a glowing review in another prominent Dutch newspaper.

In early 2002, Dijkstra returned to Nuenen, incurably ill with cancer. The news spread quickly in the computer science community and was invariably met with deep sadness. The last time I saw Dijkstra was at his home in July 2002. As was usually the case with visitors, he collected me by car from the nearest train station a couple of kilometers away from his house. During the visit, we spoke together, shared lunch, and he told me that he did not have much time left. He also gave me a copy of EWD1318, "Coxeter's Rabbit," dated April 14, 2002, mentioning that it would be his final report.[47]

Dijkstra passed away a month later. His funeral was attended by a number of his colleagues, including several from Austin. In his eulogy, Hoare reflected on Dijkstra's immense contributions to the development of his field:

> He would lay the foundations that would establish computing as a rigorous scientific discipline; and in his research and in his teaching and in his writing, he would pursue perfection to the exclusion of all other concerns. From these commitments he never deviated, and that is how he has made to his chosen subject of study the greatest contribution that any one person could make in any one lifetime.[48]

J Strother Moore, then chairman of the Computer Science Department in Austin, also spoke warmly and evocatively of Dijkstra: "He was like a man with a light in the darkness. He illuminated virtually every issue he discussed."[49]

Obituaries subsequently were published in several leading newspapers, including *The New York Times*, *The Washington Post*, and *The Guardian*. Extended commemorative pieces and reminiscences appeared during the months that followed, in which Dijkstra was variously acclaimed as a pioneer, prophet, sage, and genius.[50]

D IJKSTRA'S ENDURING influence in computer science is not confined solely to his research. He held strong opinions about many aspects of the field, most notably about programming languages and the teaching of programming, but also the purpose of education and research.

Dijkstra did not shy away from controversies. He was a dedicated contrarian who reveled in expressing extreme and unconventional opinions. I saw this firsthand in 1977 during a large computer science conference in Toronto. Audiences for plenary lectures at the event numbered somewhere between several hundred and a thousand attendees. Each of the lectures concluded with a few polite audience questions for the speaker, generally a leading expert in his area. I have a vivid recollection of Dijkstra standing up at the end of one lecture and delivering a stinging rebuke to the speaker. Contrary to appearances, he was hoping to provoke an informative discussion. The chairman was visibly taken aback by Dijkstra's intervention and appeared at a loss as to how he should proceed.

Dijkstra was also unafraid to voice harsh critiques at smaller gatherings. In the late 1970s, I attended a seminar in Utrecht with about twenty other participants, including Dijkstra. He repeatedly interrupted a highly respected lecturer to query him about his use of terminology and abbreviations. Halfway through the presentation Dijkstra abruptly got up and left. Other stories in a similar vein were far from uncommon and circulated throughout the field.

Dijkstra took his work as a reviewer extremely seriously and his reports were detailed and carefully thought out. Some of these reviews were undertaken at his own initiative and appeared as EWDs. These included a positive review of a 400-page computer science book that had no obvious connection to his research.[51] He also produced extensive and thoughtful comments on manuscripts sent to him by his colleagues who had adopted his notation or methodology, or whose research he deemed important.

In 1977, Dijkstra wrote a vitriolic review of a report, "Social Processes and Proofs of Theorems and Programs," by Richard De Millo, Richard Lipton, and Alan Perlis. The report later appeared as a journal paper.[52] Dijkstra distributed his review as EWD638, "A Political Pamphlet from the Middle Ages," in which he referred to the report as "a very ugly paper."[53] The authors had argued that "program verification is bound to fail," a view Dijkstra vehemently disagreed with.

Some of these reviews led to further correspondence with the original authors. In 1978, Dijkstra distributed a detailed and scathing review of the 1977 Turing Award Lecture delivered by John Backus. In EWD692 he argued that the lecture "suffers badly from aggressive oversell-ing."[54] At the time, Backus was one of the most prominent working computer scientists. He was the co-inventor of a standard notation used to describe the syntax of programming languages, known as Backus–Naur form, and had led the team that designed and implemented Fortran, the first high-level programming language. In his lecture,

Backus had advocated for an alternative style of programming, known as functional programming. Four years ago, Jiahao Chen discovered an extensive and highly critical correspondence between Backus and Dijkstra that took place following the review.[55] Dijkstra wisely kept these exchanges away from the public eye.

In some quarters, Dijkstra was viewed as arrogant and his opinions considered extreme. When cataloguing their correspondence in his papers, Backus added the comment: "This guy's arrogance takes your breath away."[56] For many, especially those who adopted his notation, Dijkstra became a figure akin to a guru. A small group of his followers even started their own EWD-like reports, all consecutively numbered and written by hand. Dijkstra seemed indifferent to such displays. "But he takes himself so seriously," a Turing Award winner once confided to me. Indeed, one sometimes had the impression that he carried the weight of computer science on his shoulders.

In 1984, I was invited to be a lecturer at the annual Marktoberdorf School, co-organized by Dijkstra. Although I regarded this as a great honor, I could not help but feel anxious about having him assessing my work. His review appeared in EWD895, "Trip report E. W. Dijkstra, Mark-toberdorf, 30 July–12 Aug. 1984." I was greatly relieved to discover his comments were not only fairly mild, but even reasonably positive in comparison to his other reviews: "Apt's lectures suffered somewhat from this [i.e., talking 'exclusively about CSP']. The examples chosen to illustrate his points were a bit elaborate, but his conscious efforts to be understood were highly appreciated."[57]

Encouraged by this appraisal, I submitted one of the papers I had presented to a peer-reviewed journal. A couple of months later, the anonymous referee reports arrived. One of the reviews was unmistakably the work of Dijkstra. A detailed criticism of what he regarded as a lack of sufficiently formal arguments, combined with a long list of demands and questions, made attempting satisfactory revisions a hopeless task. Even today, I would not know how to meet these demands because the right formalism is still lacking. At the time, it was nonetheless considered a privilege to have a paper refereed by Dijkstra.

In 1989, Dijkstra presented his views on teaching computer programming in a lecture titled "On the Cruelty of Really Teaching Computer Science" during an ACM Computer Science Conference. A transcript was later circulated as EWD1036.[58] After presenting a sweeping historical survey aimed at illustrating traditional resistance toward new ideas in science, Dijkstra argued that computer programming should be taught in a radically different way. His proposal was to teach some of the elements of mathematical logic, select a small but powerful programming language, and then concentrate on the task of constructing provably correct computer programs. In his view, programs should be considered the same way as formulas, while programming should be taught as a branch of mathematics. There was no place for running programs or for testing, both of which were considered standard practice in software engineering.

The lecture and report led to an extensive debate that still makes for interesting reading. Dijkstra's report was published in *Communications of the ACM*, along with his polite but unapologetic responses to mostly negative reactions from prominent computer scientists. While he was praised for initiating a much-needed debate, Dijkstra's recommendations were deemed unrealistic and too controversial.[59]

Dijkstra often expressed his opinions using memorable turns of phrase or maxims that caught the ears of his colleagues and were widely commented upon. Here are some examples:

- Program testing can be used to show the presence of bugs, but never to show their absence.
- Computer science is no more about computers than astronomy is about telescopes.
- The question of whether machines can think is about as relevant as the question of whether submarines can swim.
- A formula is worth a thousand pictures.

In one of his EWDs, Dijkstra collected several jibes about programming languages, such as: "The use of COBOL cripples the mind; its teaching should, therefore, be regarded as a criminal offense."[60] At the time, COBOL was one of the most widely used programming languages and these comments were not warmly received.

Some of Dijkstra's opinions were unavoidably controversial and highlighted his longstanding prejudices. When I first met him in 1975 he recommended the book *Structured Programming*, but suggested that I skip the final chapter by Hoare and Ole-Johan Dahl,[61] as it dealt with object-oriented programming. Nonetheless, object-oriented programming gradually became a universally preferred way to structure a large program. But not for Dijkstra. He was still arguing against it in 1999, pointing out during a keynote address that, "For those who have wondered: I don't think object-oriented programming is a structuring paradigm that meets my standards of elegance."[62] By that time, the popular object-oriented programming language C++ was routinely taught to first-year computer science students.

Throughout his professional career, Dijkstra remained remarkably modest. He never had a secretary; he typed or wrote all his publications himself. Most were entirely his own work and even the few that listed co-authors were clearly written by Dijkstra, or in his style. After 1979, he preferred to write by hand using a Montblanc fountain pen. His writing style became so recognizable among computer scientists in the 1980s that a fellow academic, Luca Cardelli, designed a Dijkstra font for Macintosh computers.[63] Not long after it was released, Dijkstra received a letter typeset in Cardelli's font and mistakenly assumed it was handwritten. He felt tricked by the letter and was not amused. Some years later, he was able to appreciate the humor when a colleague in Austin, Bob Boyer, adopted the font for presentations during departmental meetings.

It seems Dijkstra never applied for any grants—though he did receive at least one, to employ a PhD student—nor did he bring any money, in the form of research contracts, to the institutions he worked for. He also never purchased a computer. Eventually, in the late 1980s, he was given one as a gift by a computer company, but never used it for word processing. Dijkstra did not own a TV, a VCR, or even a mobile phone. He preferred to avoid the cinema, citing an oversensitivity to visual input. By contrast, he enjoyed attending classical music concerts.

When taking part in conferences and summer schools Dijkstra often felt uncomfortable in large groups. Unaccustomed to small talk, he usually remained awkwardly silent. Away from the work environment, however, he was completely different. From his time in Austin, I and others recall him as friendly, helpful, and eager to drop by with his wife for a short social visit that often led to engaging conversations. He and his wife liked to invite guests over, for whom he occasionally played short pieces of classical music on his Bösendorfer piano. His favorite composer was Mozart. A striking feature of Dijkstra's living room was a lectern with a large copy of the *Oxford English Dictionary*, which he found indispensable in his work. He is, incidentally, mentioned in the same dictionary in connection with the use of the words *vector* and *stack* in computing.

In Austin, Dijkstra stayed away from university politics. He was highly respected by colleagues, not least because of his collegial attitude during departmental meetings. He took his teaching duties very seriously. Exams were always oral and could last a couple of hours. Upon completion of the exam, an informal chat followed during which the student was presented with a signed photo of Dijkstra and a beer—age permitting.[64] He held his weekly seminars in his office and served coffee to the students in attendance, often surprising them with his unassuming behavior.

Throughout his life, Dijkstra was an uncompromising perfectionist, always focused on tapping his creativity, unwilling to lower his standards, and indifferent toward alternative points of view. He also found it difficult to browse articles in his field to form an idea of their contents and seemed uninterested in tracking down and studying the relevant literature. For the most part, he followed the recommendations of his close colleagues and only studied the papers they suggested. As a result, his own papers often had very few, if any, bibliographic entries. The preface of *A Discipline of Programming* concludes with a frank admission: "For the absence of a bibliography I offer neither explanation nor apology." This cavalier approach led to occasional complaints from colleagues who found that their work was ignored.

Instead, Dijkstra preferred to study classic texts, such as Eric Temple Bell's *Men of Mathematics*, which he referred to occasionally during his courses in Austin, and Linus Pauling's *General Chemistry*, a book he praised in highest terms.[65] This attitude served him well during the 1960s and 1970s, but it became increasingly impractical as computer science matured.

Published in 1990, Dijkstra's *Predicate Calculus and Program Semantics*, co-written with Scholten, was a case in point. The book not only lacked references, but also exhibited a complete disregard for the work of logicians. Egon Börger penned an extensive and devastating review, claiming the approach outlined by the authors was not in any way novel, nor did it offer any advantages.[66] He also vigorously criticized the book's rudimentary history of predicate logic, in which the authors drew a line from the work of Gottfried Leibniz to that of George Boole and then to their own contributions, neglecting to mention anyone else.

In response to Börger's review, some colleagues tried to help by publishing papers that provided a useful logical assessment and clarification of Dijkstra and Scholten's approach based on their so-called calculational proofs. Dijkstra remained unrepentant. "I never felt obliged to placate the logicians," he remarked some years later in EWD1227. "If however, [logicians] only get infuriated because I don't play my game according to their rules," he added, referring specifically to Börger's review, "I cannot resist the temptation to ignore their fury and to shrug my shoulders in the most polite manner."

Dijkstra's sense of humor was, at turns, wry and terse. I once asked him how many PhD students he had. "Two," he replied, before adding, "Einstein had none."[67] On another occasion, he wrote to me that "[redacted] strengthened the Department by leaving it." In Austin, together with his wife, he purchased a Volkswagen bus, dubbed the Turing Machine, which they used to explore national parks.[68]

Dijkstra was also extremely honest. He was always insistent, for example, that the first solution to the mutual exclusion problem was found by his colleague Th. J. Dekker. In EWD1308, he admitted that F. E. J. Kruseman Aretz "still found and repaired a number of errors [in the ALGOL 60 compiler] after I had left the Mathematical Centre in 1962," and that the phrase "considered harmful" was, in fact, invented by an editor of the *Communications of the ACM*, Niklaus Wirth. Dijkstra's contribution was originally titled "A Case against the GO TO Statement." In the same EWD, he also admitted completely missing the significance of Floyd and Hoare's initial contributions to program verification. "I was really slow" he lamented.[69]

D IJKSTRA LEFT BEHIND a remarkable array of notions and concepts that have withstood the test of time: the display, the mutual exclusion problem, the semaphore, deadly embrace, the banker's algorithm, the sleeping barber and the dining philosophers problems, self-stabilization, weakest precondition, guard, and structured programming.

His shortest path algorithm is taught to all students of computer science and operations research and is always referred to as Dijkstra's algorithm. Several years ago I saw it illustrated, under this name, by means of an interactive gadget with lights and buttons at the Science Centre Singapore.

Dijkstra and Zonneveld's ALGOL 60 compiler is rightly recognized as a milestone in the history of computer science—albeit more so in Europe than elsewhere. An entire PhD thesis was recently devoted to its reconstruction and a detailed analysis.[70] The layered design of the THE multiprogramming system is discussed in several textbooks on operating systems and influenced the design of some later operating systems.

Among the concepts invented by Dijkstra, some have been reflected in book titles. An early example is *Structured Programming*, published in 1972.[71] There are now several books called *Structured Programming Using Language* X. In 1986, a book appeared with the title *Algorithms for Mutual Exclusion*—others with a similar title eventually followed—and in 2000 a book titled *Self-Stabilization* was published.[72]

Dijkstra's approaches to nondeterminism and parallelism are part of standard courses on these subjects. His proposal for a first synchronization mechanism triggered research that, in turn, resulted in the development of high-level synchronization mechanisms that are indispensable for parallel programming. His classic problems, such as the sleeping barber and dining philosophers, the latter named by Hoare, have become standard benchmarks. When methods were developed to formally verify parallel programs, the first examples tackled were Dijkstra's programs.

Yet Dijkstra's most enduring contribution may well be indirect—in software engineering. The challenge of producing correct software was an ongoing concern throughout his scientific career. In 1962, he wrote a paper, "Some Meditations on Advanced Programming," in which he raised the issue of program correctness and expressed the hope that this aspect might someday be referred to as the science of programming.[73] At a major conference in 1968, it was recognized that the availability of increasingly powerful computers led to increasingly complex and unreliable software systems, a problem termed "the software crisis." Dijkstra, who was in attendance, could not have agreed more.

In the years that followed, he produced a number of engaging and influential essays on software development in which he explicitly referred to the software crisis as an urgent problem. He forcefully argued that software systems should be built on sound design principles, and that correctness should be a driving principle behind program construction. In particular, he introduced the often-cited separation-of-concerns design principle, which, he remarked, "even if not perfectly possible, is yet the only available technique for effective ordering of one's thoughts, that I know of."[74] Following the example of Hoare and Wirth, he also advocated for various forms of abstraction and the use of assertions to annotate programs. At a later stage, he argued that the programming process itself should be viewed as a mathematical activity.

Although Dijkstra's idealized vision that programs should be constructed together with their correctness proofs has not been realized, it undoubtedly provided the impetus for new methods of structuring and developing programs—including, somewhat paradoxically, the emergence of the object-oriented programming paradigm that he so vigorously opposed. This vision also helped motivate the design of new programming languages, along with platforms and systems to facilitate the programming process. Finally, Dijkstra's "Notes on Structured Programming" was highly influential in the development of better designed and more systematic courses on programming, occasionally with an emphasis on systematic program construction and correctness.

It is difficult to find another scientist who left such an impressive mark in the history of computer science.[75]

*Krzysztof Apt is a Fellow at the Center for Mathematics and Computer Science in Amsterdam and Professor Emeritus at the University of Amsterdam.*